\documentclass[prd,showpacs,floatfix,amsmath,amssymb,nofootinbib,twocolumn]{revtex4-1}
\usepackage{graphicx,color,dcolumn,booktabs,bm}
\usepackage{longtable,comment,braket}
\usepackage{times}
\usepackage{overpic}
\usepackage{amssymb,tipa}
\usepackage{indentfirst}
\usepackage{feynmf}   
\usepackage{slashed}  
\usepackage{cases}
\usepackage{epstopdf}
\usepackage{psfrag}
\usepackage{subfigure}
\usepackage{color}
\usepackage{multirow}

\usepackage[colorlinks, citecolor=blue,anchorcolor=red,menucolor=red,linkcolor=blue,filecolor=red,runcolor=red,urlcolor=blue,frenchlinks=red]{hyperref}

\usepackage{ulem}
\def\be{\begin{equation}}
\def\ee{\end{equation}}

\begin{document}
\title{Spectrum of the strange hidden charm molecular pentaquarks in chiral effective field theory}
\author{Bo Wang$^{1,2}$}\email{bo-wang@pku.edu.cn}
\author{Lu Meng$^{2}$}\email{lmeng@pku.edu.cn}
\author{Shi-Lin Zhu$^{2,1}$}\email{zhusl@pku.edu.cn}
\affiliation{
$^1$Center of High Energy Physics, Peking University, Beijing 100871, China\\
$^2$ School of Physics and State Key Laboratory of Nuclear Physics
and Technology, Peking University, Beijing 100871, China}

\begin{abstract}
We calculate the effective potentials of the
$\Xi_c\bar{D}^{(\ast)}$, $\Xi_c^\prime\bar{D}^{(\ast)}$ and
$\Xi_c^\ast\bar{D}^{(\ast)}$ systems with the chiral effective field
theory up to the next-to-leading order. We simultaneously consider
the short-, intermediate- and long-range interactions. With the
newly observed $P_c$ spectra as inputs, we construct the quark-level
contact Lagrangians to relate the low energy constants to those of
$\Sigma_c\bar{D}^{(\ast)}$ with the help of quark model. Our
calculation indicates there are seven bound states in the $I=0$
strange hidden charm
$[\Xi_c^\prime\bar{D}^{(\ast)}]_J~(J=\frac{1}{2},\frac{3}{2})$ and
$[\Xi_c^\ast\bar{D}^{(\ast)}]_J~(J=\frac{1}{2},\frac{3}{2},\frac{5}{2})$
systems. Our analyses also disfavor the $\Lambda_c\bar{D}^{(\ast)}$
bound states. However, we obtain three new hadronic molecules in the
isoscalar $[\Xi_c\bar{D}^{(\ast)}]_J~(J=\frac{1}{2},\frac{3}{2})$
systems. The masses of $[\Xi_c\bar{D}]_{1/2}$,
$[\Xi_c\bar{D}^{\ast}]_{1/2}$ and $[\Xi_c\bar{D}^{\ast}]_{3/2}$ are
predicted to be $4319.4^{+2.8}_{-3.0}$ MeV, $4456.9^{+3.2}_{-3.3}$
MeV and $4463.0^{+2.8}_{-3.0}$ MeV, respectively. We also notice the
one-eta-exchange influence is rather feeble. Binding solutions in
the $I=1$ channels are nonexistent. We hope the future analyses at
LHCb can seek for these new $P_{cs}$s in the $J\psi\Lambda$ final
states, especially near the thresholds of $\Xi_c\bar{D}^{(\ast)}$.
\end{abstract}
 \maketitle

\section{Introduction}\label{Introduction}

In the past decades, the renaissance of hadron physics was
witnessed. Mesons and baryons, which have the internal
configurations $q\bar{q}$ and $qqq$ respectively, have been
extensively studied with lattice QCD and various QCD inspired
models. The abundant conventional hadrons in the Reviews of Particle
Physics~\cite{Tanabashi:2018oca} reflect the great victory of the
quark model. Other more complicated quark configurations, such as
$qq\bar{q}\bar{q}$, $q\bar{q}qqq$, and $qqq\bar{q}\bar{q}\bar{q}$,
etc., are not forbidden by QCD. Thus hunting for these type states
is a long standing problem for theorists and experimenters.
$X(3872)$ is the poster child that opened a new era for hadron
physics~\cite{Choi:2003ue}. After that, more and more $XYZ$ states
were discovered. The multiquark matter becomes one of the hottest
topics in recent
years~\cite{Chen:2016qju,Guo:2017jvc,Liu:2019zoy,Lebed:2016hpi,Esposito:2016noz,Brambilla:2019esw}.

Very recently, the LHCb Collaboration reported the observation of
three pentaquark states $P_c(4312)$, $P_c(4440)$ and
$P_c(4457)$~\cite{Aaij:2019vzc}. Their masses lie several to tens
MeV below the $\Sigma_c\bar{D}$ and $\Sigma_c\bar{D}^\ast$
thresholds, thus the molecular explanation is naturally proposed in
many
works~\cite{Meng:2019ilv,Wang:2019ato,Liu:2019tjn,Chen:2019asm,Xiao:2019aya,He:2019ify,Xiao:2019mst,Voloshin:2019aut,Guo:2019fdo,Guo:2019kdc,Burns:2019iih,Wang:2019spc}.
The $J^P$ quantum numbers are undetermined yet, but the
theoretically favored ones in the molecular scenario for these three
$P_c$s are $\frac{1}{2}^-$, $\frac{1}{2}^-$ and $\frac{3}{2}^-$,
respectively. Therefore, if the forthcoming measurements for the
$J^P$s indeed meet the predictions from the molecular pictures,
which would give more robust support for the molecular
interpretations.

The $XYZ$ and $P_c$ states are the candidates of the hidden-charm
multiquark states with inner quark components $Q\bar{Q}q\bar{q}$ and
$Q\bar{Q}qqq$, respectively. As indicated in
Refs.~\cite{Lee:2011rka,Li:2012bt,Li:2014gra}, it seems the heavy
quark core plays an important role in stabilizing the exotic
clusters. This is indeed the case in the atomic physics. For
example, the hydrogen molecule consists of two protons and two
electrons, which stably exists in the nature. In the hadronic
molecular scenario, the interaction between two color singlets
(e.g., $\Sigma_c$ and $\bar{D}$) is very similar to that between
electroneutral atoms (e.g., $H$ and $H$). The ``covalent bond" in
the former is attributed to the residual strong interactions, which
is equivalently described by the pion-exchange in chiral effective
theory or the meson-exchange (e.g., $\pi$, $\rho$, $\sigma$, ...) in
one-boson-exchange model. Therefore, one could actually anticipate
the existence of more hadronic molecules in the charmed
baryon-anticharmed meson systems when the flavor symmetry group is
enlarged to SU(3).

Starting from the deuteron (an $I=0$ loosely bound $np$ molecule),
one can notice that the interactions between two heavy matter fields
tend to form the bound states in the lowest isospin channels.
$X(3872)$ is another example, which is a good candidate of the
$D^0\bar{D}^{\ast0}$ molecule with isospin $I=0$. The newly reported
$P_c$s are widely accepted as the $\Sigma_c\bar{D}^{(\ast)}$ bound
states with isospins $I=\frac{1}{2}$. Some investigations on the
$DD^\ast$ and $\bar{B}^{(\ast)}\bar{B}^{(\ast)}$ systems also
demonstrated the existence of bound states in $I=0$
channels~\cite{Xu:2017tsr,Wang:2018atz}.

In our previous work~\cite{Wang:2019ato}, we have systematically
investigated the interactions of the $\Sigma_c\bar{D}$,
$\Sigma_c\bar{D}^\ast$, $\Sigma_c^\ast\bar{D}$ and
$\Sigma_c^\ast\bar{D}^\ast$ systems in chiral effective field
theory. We simultaneously reproduced the newly observed three $P_c$s
as the $I=\frac{1}{2}$ hidden-charm $\Sigma_c\bar{D}$ and
$\Sigma_c\bar{D}^\ast$ molecules by introducing the $\Lambda_c$
contribution in the two-pion-exchange loop diagrams. In this work,
we extend our study to the $\Xi_c\bar{D}^{(\ast)}$,
$\Xi_c^\prime\bar{D}^{(\ast)}$ and $\Xi_c^\ast\bar{D}^{(\ast)}$
systems to see whether there exist the bound states in the lowest
isospin, i.e., $I=0$ channels. Likewise, these strange hidden charm
molecular states might be reconstructed in the $J/\psi\Lambda$
channel at the LHCb experiment. Some investigations suggest
searching for these states in the decay modes $\Lambda_b(\Xi_b)\to
J/\psi\Lambda
K(\eta)$~\cite{Lu:2016roh,Feijoo:2015kts,Chen:2015sxa}.

Based on Ref.~\cite{Wang:2019ato}, we further study the effective
potentials of six systems, i.e., $\Xi_c\bar{D}^{(\ast)}$,
$\Xi_c^\prime\bar{D}^{(\ast)}$ and $\Xi_c^\ast\bar{D}^{(\ast)}$.
They all contain one strange quark. The short-range contact
interaction, long-range one-pion-exchange contribution and
intermediate-range two-pion-exchange diagrams are all included in
the framework of chiral effective field theory (For the reviews of
chiral effective field theory, we refer
to~\cite{Bernard:1995dp,Epelbaum:2008ga,Machleidt:2011zz,Meissner:2015wva,Hammer:2019poc}).
Considering the hadronic molecules are shallowly bound states, the
strange quark dynamics are freezed in the present calculations,
which contribution is partially involved in the contact terms. We
ignore the $\eta$ and $K$ meson contributions in the loops. The low
energy constants (LECs) are well determined by fitting the $P_c$
spectra. In this way, we predict the possible strange hidden charm
molecular pentaquarks.

This paper is organized as follows. In Sec.~\ref{Lagrangians}, we
give the effective Lagrangians constructed with the chiral and heavy
quark symmetries. In Sec.~\ref{EffecctivePotential}, we show the
expressions of effective potentials. In Sec.~\ref{NumericalResults},
we give the numerical results and discussions. In
Sec.~\ref{Summary}, we conclude this work with a short summary. In
Appendix~\ref{app:QM}, we bridge the LECs to those of
$\Sigma_c^{(*)}\bar{D}^{(*)}$ with quark model.

\section{Effective Lagrangians with the chiral and heavy quark symmetries}\label{Lagrangians}

The effective Lagrangians can be classified as the pion and contact
interactions, respectively. We first show the Lagrangians of the
charmed baryon (anticharmed mesons) and light pseudoscalar meson
interaction. For the charmed baryons, the matrix forms of the
spin-${1}\over{2}$ antitriplet, spin-${1}\over{2}$ and
spin-${3}\over{2}$ sextets in the SU(3) flavor space are expressed
as
\begin{widetext}
\begin{eqnarray}
\mathcal{B}_{\bar{3}}=\left[ \begin{array}{ccc}
0&\Lambda_c^+&\Xi_c^+\\
-\Lambda_c^+&0&\Xi_c^0\\
-\Xi_c^+&-\Xi_c^0&0
\end{array} \right],\quad\quad
\mathcal{B}_6=\left[ \begin{array}{ccc}
\Sigma_c^{++}&\frac{\Sigma_c^+}{\sqrt{2}}&\frac{\Xi_c^{\prime+}}{\sqrt{2}}\\
\frac{\Sigma_c^+}{\sqrt{2}}&\Sigma_c^0&\frac{\Xi_c^{\prime0}}{\sqrt{2}}\\
\frac{\Xi_c^{\prime+}}{\sqrt{2}}&\frac{\Xi_c^{\prime0}}{\sqrt{2}}&\Omega_c^0
\end{array} \right],\quad\quad
\mathcal{B}_{6}^{\ast\mu}=\left[ \begin{array}{ccc}
\Sigma_c^{\ast++}&\frac{\Sigma_c^{\ast+}}{\sqrt{2}}&\frac{\Xi_c^{\ast+}}{\sqrt{2}}\\
\frac{\Sigma_c^{\ast+}}{\sqrt{2}}&\Sigma_c^{\ast0}&\frac{\Xi_c^{\ast0}}{\sqrt{2}}\\
\frac{\Xi_c^{\ast+}}{\sqrt{2}}&\frac{\Xi_c^{\ast0}}{\sqrt{2}}&\Omega_c^{\ast0}
\end{array} \right]^\mu.
\end{eqnarray}
The leading order chiral Lagrangians for the charmed baryons and
light pseudoscalar mesons in the super-field notation
read~\cite{Cho:1992cf,Cheng:1993kp,Cho:1992nt,Cho:1992gg}
\begin{eqnarray}\label{Baryon_Lag_SF}
\mathcal{L}_{\mathcal{B}\varphi}&=&-\mathrm{Tr}\left[\bar{\psi}^\mu iv\cdot D\psi_\mu\right]+ig_a\epsilon_{\mu\nu\rho\sigma}\mathrm{Tr}\left[\bar{\psi}^\mu u^\rho v^\sigma\psi^\nu\right]+i\frac{\delta_a}{2}\mathrm{Tr}\left[\bar{\psi}^\mu\sigma_{\mu\nu}\psi^\nu\right]\nonumber\\
&&+\frac{1}{2}\mathrm{Tr}\left[\bar{\mathcal{B}}_{\bar{3}}(iv\cdot
D)\mathcal{B}_{\bar{3}}\right]+g_b\mathrm{Tr}\left[\bar{\psi}^\mu
u_\mu\mathcal{B}_{\bar{3}}+\mathrm{H.c.}\right],
\end{eqnarray}
where $\mathrm{Tr}[\cdots]$ denotes the trace in flavor space. The
covariant derivative
$D_\mu\psi=\partial_\mu\psi+\Gamma_\mu\psi+\psi\Gamma_\mu^T$.
$g_a\simeq-1.47$ and $g_b\simeq1.04$ are the axial
couplings~\cite{Jiang:2015xqa,Meng:2018gan,Wang:2018cre}.
$\delta_a=m_{\Xi_c^\ast}-m_{\Xi_c^\prime}$ is the mass splitting
between $\Xi_c^\ast$ and $\Xi_c^\prime$ in this calculation.
$v^\mu=(1,\mathbf{0})$ represents the four-velocity of a slowly
moving heavy baryon. The super-fields $\psi^\mu$ and
$\bar{\psi}^\mu$ are defined as
\begin{eqnarray}
\psi^\mu=\mathcal{B}_{6}^{\ast\mu}-\frac{1}{\sqrt{3}}(\gamma^\mu+v^\mu)\gamma^5\mathcal{B}_6,\quad\quad\quad\bar{\psi}^\mu=\bar{\mathcal{B}}_{6}^{\ast\mu}+\frac{1}{\sqrt{3}}\bar{\mathcal{B}}_6\gamma^5(\gamma^\mu+v^\mu).
\end{eqnarray}
In addition, the chiral connection $\Gamma^\mu$ and axial-vector
current $u^\mu$ read respectively
\begin{eqnarray}
\Gamma_\mu&\equiv&\frac{1}{2}\left[\xi^\dag,\partial_\mu
\xi\right],\quad\quad\quad
u_\mu\equiv\frac{i}{2}\left\{\xi^\dag,\partial_\mu \xi\right\},
\end{eqnarray}
where
\begin{eqnarray}\label{pifield}
\xi^2=U=\exp\left(\frac{i\varphi}{f_\pi}\right),\quad\quad\quad
\varphi=\left[ \begin{array}{ccc}
\pi^0+\frac{\eta}{\sqrt{3}}&\sqrt{2}\pi^+&\sqrt{2}K^+\\
\sqrt{2}\pi^-&-\pi^0+\frac{\eta}{\sqrt{3}}&\sqrt{2}K^0\\
\sqrt{2}K^0&\sqrt{2}\bar{K}^0&-\frac{2\eta}{\sqrt{3}}
\end{array} \right],
\end{eqnarray}
and $f_\pi=92.4$ MeV is the pion decay constant. Expanding
Eq.~\eqref{Baryon_Lag_SF} one can get the coupling terms among
$\mathcal{B}_{\bar{3}}$, $\mathcal{B}_6$ and
$\mathcal{B}_{6}^{\ast\mu}$. The detailed forms and the
corresponding axial couplings can be found in
Refs.~\cite{Wang:2019ato,Jiang:2015xqa,Meng:2018gan,Wang:2018cre}.
\end{widetext}

The leading order Lagrangians that delineate the interactions
between the anticharmed mesons and light Goldstones
read~\cite{Wise:1992hn,Manohar:2000dt}
\begin{eqnarray}\label{Meson_Lag_SF}
\mathcal{L}_{\tilde{\mathcal{H}}\varphi}=-i\langle\bar{\tilde{\mathcal{H}}}
v \cdot \mathcal{D}
\tilde{\mathcal{H}}\rangle-\frac{1}{8}\delta_b\langle\bar{\tilde{\mathcal{H}}}
\sigma^{\mu \nu} \tilde{\mathcal{H}}\sigma_{\mu \nu}\rangle+
g\langle\bar{\tilde{\mathcal{H}}} \slashed{u}\gamma_{5}
\tilde{\mathcal{H}}\rangle,\nonumber\\
\end{eqnarray}
where $\langle\cdots\rangle$ represents the trace in spinor space.
The covariant derivative $\mathcal{D}_\mu=\partial_\mu+\Gamma_\mu$.
$\delta_b$ is defined as $\delta_b=m_{\bar{D}^\ast}-m_{\bar{D}}$.
$g\simeq-0.59$ stands for the axial coupling, which can be extracted
from the partial decay width of $D^{\ast+}\to
D^0\pi^+$~\cite{Tanabashi:2018oca}. The $\tilde{\mathcal{H}}$ is the
super-field for the anticharmed mesons, which reads
\begin{eqnarray}
\tilde{\mathcal{H}}=\left[\tilde{P}_{\mu}^{*} \gamma^{\mu}+i
\tilde{P} \gamma_{5}\right]
\frac{1-\slashed{v}}{2},\bar{\tilde{\mathcal{H}}}=\frac{1-\slashed{v}}{2}\left[\tilde{P}_{\mu}^{*
\dagger}\gamma^{\mu}+i \tilde{P}^{\dagger}
\gamma_{5}\right],\nonumber
\end{eqnarray}
with $\tilde{P}=(\bar{D}^0,D^-,D_s^-)^T$ and
$\tilde{P}^\ast=(\bar{D}^{\ast0},D^{\ast-},D^{\ast-}_s)^T$,
respectively.

The contact Lagrangians that describe the short distance
interactions between the charmed baryon sextets and anticharmed
mesons have been constructed in
Refs.~\cite{Meng:2019ilv,Wang:2019ato} with the super-field
representations, which read
\begin{widetext}
\begin{eqnarray}\label{Contact_Lag_BM}
\mathcal{L}_{\tilde{\mathcal{H}}\mathcal{B}_6^{(\ast)}}&=&D_a \langle\bar{\tilde{\mathcal{H}}}\tilde{\mathcal{H}}\rangle\mathrm{Tr}\big[\bar{\psi}^\mu \psi_\mu\big]+iD_b\epsilon_{\sigma\mu\nu\rho}v^\sigma\langle\bar{\tilde{\mathcal{H}}}\gamma^\rho\gamma_5\tilde{\mathcal{H}}\rangle\mathrm{Tr}\left[\bar{\psi}^\mu \psi^\nu\right]\nonumber\\
&&+E_a
\langle\bar{\tilde{\mathcal{H}}}\lambda^i\tilde{\mathcal{H}}\rangle\mathrm{Tr}\big[\bar{\psi}^\mu\lambda_i
\psi_\mu\big]+iE_b\epsilon_{\sigma\mu\nu\rho}v^\sigma\langle\bar{\tilde{\mathcal{H}}}\gamma^\rho\gamma_5\lambda^i\tilde{\mathcal{H}}\rangle\mathrm{Tr}\big[\bar{\psi}^\mu\lambda_i
\psi^\nu\big],
\end{eqnarray}
where $D_a$, $D_b$, $E_a$ and $E_b$ are the LECs. They can be
determined by fitting the $P_c$ spectra. $\lambda_i$ denotes the
Gell-Mann matrices.

Besides, we also need the Lagrangians to depict the contact
interactions of the charmed baryon antitriplet and anticharmed
mesons. They can be analogously constructed as follows,
\begin{eqnarray}\label{Contact_Lag_B3M}
\mathcal{L}_{\tilde{\mathcal{H}}\mathcal{B}_{\bar{3}}}&=&\tilde{D}_a \langle\bar{\tilde{\mathcal{H}}}\tilde{\mathcal{H}}\rangle\mathrm{Tr}\big[\bar{\mathcal{B}}_{\bar{3}} \mathcal{B}_{\bar{3}}\big]+\tilde{D}_b\langle\bar{\tilde{\mathcal{H}}}\gamma^\rho\gamma_5\tilde{\mathcal{H}}\rangle\mathrm{Tr}[\bar{\mathcal{B}}_{\bar{3}}\gamma_\rho\gamma_5 \mathcal{B}_{\bar{3}}]\nonumber\\
&&+\tilde{E}_a
\langle\bar{\tilde{\mathcal{H}}}\lambda^i\tilde{\mathcal{H}}\rangle\mathrm{Tr}[\bar{\mathcal{B}}_{\bar{3}}\lambda_i
\mathcal{B}_{\bar{3}}]+\tilde{E}_b\langle\bar{\tilde{\mathcal{H}}}\gamma^\rho\gamma_5\lambda^i\tilde{\mathcal{H}}\rangle\mathrm{Tr}[\bar{\mathcal{B}}_{\bar{3}}\gamma_\rho\gamma_5\lambda_i
\mathcal{B}_{\bar{3}}],
\end{eqnarray}
\end{widetext}
where $\tilde{D}_a$, $\tilde{D}_b$, $\tilde{E}_a$ and $\tilde{E}_b$
are another sets of the LECs. These LECs are different from the ones
in Eq.~\eqref{Contact_Lag_BM}, since the $\mathcal{B}_{\bar{3}}$ and
$\mathcal{B}_6^{(\ast)}$ are not the partner states under heavy
quark spin symmetry. But we can establish the corresponding
relationship with the $D_a$, $D_b$, $E_a$ and $E_b$ with the help of
quark model. We show this operation in the Appendix~\ref{app:QM}.

\section{Expressions of the effective potentials}\label{EffecctivePotential}

There exists a simple relation between the effective potential and
scattering amplitude in momentum space under the Breit
approximation,
\begin{eqnarray}
\mathcal{V}(\bm{q})=-\frac{\mathcal{M}(\bm{q})}{\sqrt{2\Pi_i m_i
2\Pi_f m_f}},
\end{eqnarray}
where $m_i$ and $m_f$ are the masses of the initial and final
states. $\bm{q}$ denotes the transferred momentum between two
scattering particles. Then the effective potential in coordinate
space can be easily obtained by Fourier transformation, which yields
\begin{eqnarray}\label{Four_Tansf}
\mathcal{V}(r)=\int\frac{d^3\bm{q}}{(2\pi)^3}e^{i\bm{q\cdot
r}}\mathcal{V}(\bm q)\mathcal{F}(\bm q),
\end{eqnarray}
where a Gauss regulator
$\mathcal{F}(\bm{q})=\exp(-\bm{q}^{2n}/\Lambda^{2n})$ is introduced
to suppress the high momentum
contribution~\cite{Ordonez:1995rz,Epelbaum:1999dj}. $n=2$ is used in
this work~\cite{Epelbaum:2014efa,Entem:2003ft}. Considering the
$\rho$ meson mass $m_\rho$ is treated as the typical hard scale in
chiral effective theory, thus the cutoff $\Lambda$ should be smaller
than $m_\rho$ (A detailed discussion on the range of $\Lambda$ can
be found in Ref.~\cite{Wang:2019ato}). The $\Lambda$ is chosen to be
around $0.5$ GeV to perform fittings and give
predictions~\cite{Meng:2019ilv,Wang:2019ato,Machleidt:2011zz,Epelbaum:2014efa}.

\begin{figure*}[!hptb]
\begin{centering}
    \scalebox{1.8}{\includegraphics[width=\columnwidth]{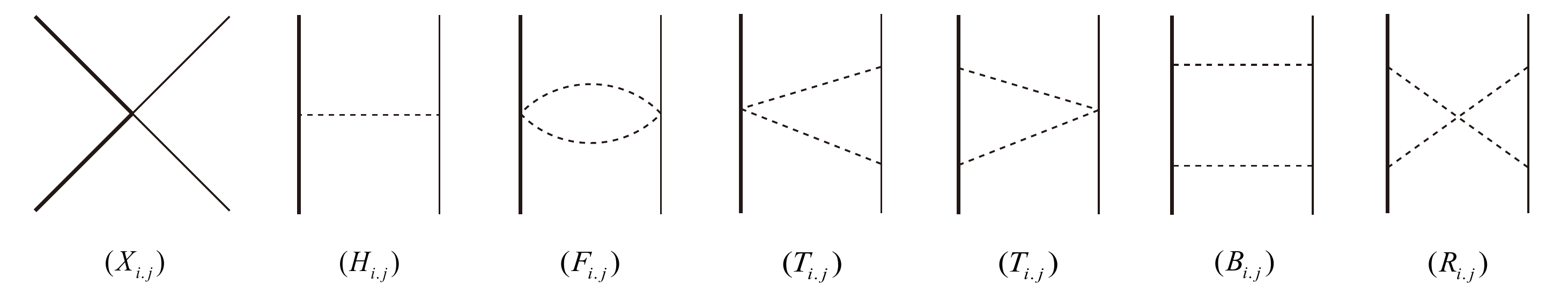}}
    \caption{Topological Feynman diagrams for the $\Xi_c\bar{D}^{(\ast)}$, $\Xi_c^\prime\bar{D}^{(\ast)}$ and $\Xi_c^\ast\bar{D}^{(\ast)}$ systems. We use the thick line, thin line and dashed line to
denote the charmed baryons, anticharmed mesons and pion,
respectively. Graphs ($X_{i.j}$) and ($H_{i.j}$) stand for the
leading order contact interaction and one-pion-exchange diagram,
respectively [Diagram ($H_{i.j}$) vanishes for the system containing
one $\bar{D}$ meson]. ($F_{i.j}$), ($T_{i.j}$), ($B_{i.j}$) and
($R_{i.j}$) represent topological structures of two-pion-exchange
diagrams, respectively.\label{TwoPion_Loop}}
\end{centering}
\end{figure*}

The topological diagrams are shown in Fig.~\ref{TwoPion_Loop}. There
are three types of Feynman diagrams in our calculations, i.e., the
leading order contact interaction, one-pion-exchange diagram, and
the next-to-leading order two-pion-exchange diagrams. The Feynman
diagrams for the $\Xi_c\bar{D}^{(\ast)}$
($\Xi_c^\prime\bar{D}^{(\ast)}$) and $\Xi_c^\ast\bar{D}^{(\ast)}$
systems are totally the same as the $\Sigma_c\bar{D}^{(\ast)}$ and
$\Sigma_c^\ast\bar{D}^{(\ast)}$, respectively. In other words, one
can build the following correspondence,
\begin{eqnarray}\label{corresp}
\Xi_c^\prime\bar{D}^{(\ast)}[\Xi_c\bar{D}^{(\ast)}]&\Leftrightarrow&\Sigma_c\bar{D}^{(\ast)},\quad\Xi_c^\ast\bar{D}^{(\ast)}\Leftrightarrow\Sigma_c^\ast\bar{D}^{(\ast)}.
\end{eqnarray}
So we do not explicitly show the detailed graphs for each systems at
each order. One can find them in figures $2$$-$$7$ of
Ref.~\cite{Wang:2019ato}.

In the following, we write down the leading order contact potential
of each system, which can be easily obtained by expanding the
Lagrangians in Eqs.~\eqref{Contact_Lag_BM}
and~\eqref{Contact_Lag_B3M}, respectively.
\begin{eqnarray}
\mathcal{V}_{\Xi_c^\prime \bar{D}}^{X_{i.j}}&=&-D_a-2\mathcal{G}E_a,\\
\mathcal{V}_{\Xi_c^\prime\bar{D}^\ast}^{X_{i.j}}&=&-D_a-2\mathcal{G}E_a-\frac{2}{3}\left(D_b+2\mathcal{G}E_b\right)\boldsymbol{\sigma}\cdot\mathbf{T},\label{PXicDast}\\
\mathcal{V}_{\Xi_c^\ast \bar{D}}^{X_{i.j}}&=&-D_a-2\mathcal{G}E_a,\\
\mathcal{V}_{\Xi_c^\ast \bar{D}^\ast}^{X_{i.j}}&=&-D_a-2\mathcal{G}E_a-\left(D_b+2\mathcal{G}E_b\right)\boldsymbol{\sigma}_{rs}\cdot\mathbf{T},\\
\mathcal{V}_{\Xi_c \bar{D}}^{X_{i.j}}&=&2\tilde{D}_a+4\mathcal{G}\tilde{E}_a,\\
\mathcal{V}_{\Xi_c\bar{D}^\ast}^{X_{i.j}}&=&2\tilde{D}_a+4\mathcal{G}\tilde{E}_a+\left(2\tilde{D}_b+4\mathcal{G}\tilde{E}_b\right)\boldsymbol{\sigma}\cdot\mathbf{T}\label{PpXicDast},
\end{eqnarray}
where $\mathcal{G}=\mathbf{I}_1\cdot\mathbf{I}_2-1/12$
[$\mathbf{I}_1$ and $\mathbf{I}_2$ are the isospin operators of
$\Xi_c^\prime$ ($\Xi_c^\ast$, $\Xi_c$) and $\bar{D}^{(\ast)}$,
respectively]. The operators $\boldsymbol{\sigma}$,
$\boldsymbol{\sigma}_{rs}$ and $\mathbf{T}$ are related to the spin
operators of the spin-$1\over{2}$ baryon, spin-$3\over{2}$ baryon
and spin-$1$ meson as $\frac{1}{2}\boldsymbol{\sigma}$,
$\frac{3}{2}\boldsymbol{\sigma}_{rs}$ and $-\mathbf{T}$,
respectively (see Ref.~\cite{Wang:2019ato} for details).

The expressions of the one-pion-exchange diagrams for
$\Xi_c^\prime\bar{D}^{\ast}$ and $\Xi_c^\ast\bar{D}^{\ast}$ are the
same as the $\Sigma_c\bar{D}^{\ast}$ and
$\Sigma_c^\ast\bar{D}^{\ast}$ in Ref.~\cite{Wang:2019ato} up to
different matrix elements of $\mathbf{I}_1\cdot\mathbf{I}_2$
operator. For $\Xi_c\bar{D}^\ast$, the coupling between $\Xi_c$ and
$\pi$ vanishes since the pion does not couple to the scalar
isoscalar light diquark within $\Xi_c$ because of the parity and
angular momentum conservation. Thus the one-pion-exchange does not
contribute to the effective potential of the $\Xi_c\bar{D}^\ast$
system.

For the two-pion-exchange diagrams, the graph ($F_{i.j}$) is
governed by the chiral connection term, thus all the systems share
one single expression, i.e.,
\begin{eqnarray}
\mathcal{V}_{\mathrm{Sys.}}^{F_{i.j}}&=&(\mathbf{I}_1\cdot\mathbf{I}_2)\frac{1}{f_\pi^4}J_{22}^F(m_\pi,q).
\end{eqnarray}
For graphs ($T_{i.j}$), ($B_{i.j}$) and ($R_{i.j}$), their
analytical expressions generally have the following structures,
\begin{widetext}
\begin{eqnarray}
\mathcal{V}_{\mathrm{Sys.}}^{T_{i.j}}&=&(\mathbf{I}_1\cdot\mathbf{I}_2)\frac{\mathcal{C}_{\mathrm{Sys.}}^{T_{i.j}}}{f_\pi^4}\left[\mathcal{C}_{1}^{T_{i.j}}J_{34}^T-\bm{q}^2\mathcal{C}_{2}^{T_{i.j}}(J_{24}^T+J_{33}^T)\right](m_\pi,\omega,q),
\end{eqnarray}
\begin{eqnarray}
\mathcal{V}_{\mathrm{Sys.}}^{B_{i.j}}&=&\left(\frac{1}{8}-\frac{1}{3}\mathbf{I}_1\cdot\mathbf{I}_2\right)\frac{\mathcal{C}_{\mathrm{Sys.}}^{B_{i.j}}}{f_\pi^4}\left[\mathcal{C}_{1}^{B_{i.j}}J_{41}^B-\bm{q}^2\mathcal{C}_{2}^{B_{i.j}}(J_{31}^B+J_{42}^B)
-\bm{q}^2\mathcal{C}_{3}^{B_{i.j}}J_{21}^B+\bm{q}^4\mathcal{C}_{4}^{B_{i.j}}(J_{22}^B+2J_{32}^B+J_{43}^B)\right](m_\pi,\omega,\delta,q),\nonumber\\
\\
\mathcal{V}_{\mathrm{Sys.}}^{R_{i.j}}&=&\left(\frac{1}{8}+\frac{1}{3}\mathbf{I}_1\cdot\mathbf{I}_2\right)\frac{\mathcal{C}_{\mathrm{Sys.}}^{R_{i.j}}}{f_\pi^4}\left[\mathcal{C}_{1}^{R_{i.j}}J_{41}^R-\bm{q}^2\mathcal{C}_{2}^{R_{i.j}}(J_{31}^R+J_{42}^R)
-\bm{q}^2\mathcal{C}_{3}^{R_{i.j}}J_{21}^R+\bm{q}^4\mathcal{C}_{4}^{R_{i.j}}(J_{22}^R+2J_{32}^R+J_{43}^R)\right](m_\pi,\omega,\delta,q),\nonumber\\
\end{eqnarray}
\end{widetext}
where the subscript ``Sys." denotes the corresponding system, such
as $\Xi_c^\prime\bar{D}^{(\ast)}$ and so on. The superscript
``$T_{i.j}$", ``$B_{i.j}$" and ``$R_{i.j}$" represent the labels of
the Feynman diagrams. Various $J$ functions, such as $J_x^{T}$,
$J_x^{B}$ and $J_x^{R}$ are the scalar loop functions, which are
defined and given in
Refs.~\cite{Meng:2019ilv,Wang:2019ato,Wang:2018atz}. The
coefficients $\mathcal{C}_i^{T_{i.j}}~(i=1,2)$ and
$\mathcal{C}_i^{B_{i.j}}(\mathcal{C}_i^{R_{i.j}})~(i=1,\dots,4)$ in
the square brackets are the same as the expressions for
$\Sigma_c^{(\ast)}\bar{D}^{(\ast)}$~\cite{Wang:2019ato}. According
to the correspondence in Eq.~\eqref{corresp}, one can easily get
them by matching with the results in Ref.~\cite{Wang:2019ato}. The
coefficients $\mathcal{C}_{\mathrm{Sys.}}^{T_{i.j}}$ and
$\mathcal{C}_{\mathrm{Sys.}}^{B_{i.j}}(\mathcal{C}_{\mathrm{Sys.}}^{R_{i.j}})$
for each system are given in Tables~\ref{Csyslabel1}
and~\ref{Csyslabel3}.

\begin{table}[htbp]
\centering 
\renewcommand{\arraystretch}{1.5}
\caption{The coefficients $\mathcal{C}_{\mathrm{Sys.}}^{T_{i.j}}$
and $\mathcal{C}_{\mathrm{Sys.}}^{B_{i.j}}$ for the systems
$\Xi_c^\prime\bar{D}$, $\Xi_c\bar{D}$ and $\Xi_c^\ast\bar{D}$,
respectively. $i=1$ for $\Xi_c^\prime\bar{D}$ ($\Xi_c\bar{D}$) and
$i=3$ for $\Xi_c^\ast\bar{D}$ ($i$ is the value in the subscript of
the diagram label), where the corresponding diagrams can be found in
figures $3$ and $6$ of Ref.~\cite{Wang:2019ato}, respectively. The
unlisted coefficients $\mathcal{C}_{\mathrm{Sys.}}^{R_{i.j}}$ can be
obtained with
$\mathcal{C}_{\mathrm{Sys.}}^{R_{i.j}}=\mathcal{C}_{\mathrm{Sys.}}^{B_{i.j}}$.
Some expressions for $\mathcal{C}_{\mathrm{Sys.}}^{\mathrm{label}}$
are denoted as $X|Y$, where $X$ and $Y$ represent the intermediate
baryon in the loop among and beyond the heavy quark spin multiplet,
respectively. \label{Csyslabel1}} \setlength{\tabcolsep}{1.75mm} {
\begin{tabular}{c|ccccc}
\hline\hline
$\mathcal{C}_{\mathrm{Sys.}}^{\mathrm{label}}$& $T_{i.1}$ &$T_{i.2}$&$T_{i.3}$&$B_{i.1}$&$B_{i.2}$\\
\hline
$\Xi_c^\prime\bar{D}$&$g^2$&$\frac{1}{4}g_3^2$&$\frac{1}{4}g_1^2\big|\frac{1}{2}g_2^2$&$\frac{3}{8}g^2g_1^2\big|\frac{3}{4}g^2g_2^2$&$\frac{3}{8}g^2g_3^2$\\
$\Xi_c\bar{D}$&$g^2$&$\frac{1}{2}g_4^2$&$g_6^2\big|\frac{1}{2}g_2^2$&$\frac{3}{2}g^2g_6^2\big|\frac{3}{4}g^2g_2^2$&$\frac{3}{4}g^2g_4^2$\\
$\Xi_c^\ast\bar{D}$&$g^2$&$\frac{1}{4}g_5^2$&$\frac{1}{4}g_3^2\big|\frac{1}{2}g_4^2$&$\frac{3}{8}g^2g_5^2$&$\frac{3}{8}g^2g_3^2\big|\frac{3}{4}g^2g_4^2$\\
\hline\hline
\end{tabular}
}
\end{table}
\begin{table*}[htbp]
\centering 
\renewcommand{\arraystretch}{1.5}
\caption{The coefficients $\mathcal{C}_{\mathrm{Sys.}}^{T_{i.j}}$
and $\mathcal{C}_{\mathrm{Sys.}}^{B_{i.j}}$ for the systems
$\Xi_c^\prime\bar{D}^\ast$, $\Xi_c\bar{D}^\ast$ and
$\Xi_c^\ast\bar{D}^\ast$, respectively. $i=2$ for
$\Xi_c^\prime\bar{D}^\ast$ ($\Xi_c\bar{D}^\ast$) and $i=4$ for
$\Xi_c^\ast\bar{D}^\ast$ ($i$ is the value in the subscript of the
diagram label), where the corresponding diagrams can be found in
figures $5$ and $7$ of Ref.~\cite{Wang:2019ato}, respectively. The
unlisted coefficients $\mathcal{C}_{\mathrm{Sys.}}^{R_{i.j}}$ can be
obtained with
$\mathcal{C}_{\mathrm{Sys.}}^{R_{i.j}}=\mathcal{C}_{\mathrm{Sys.}}^{B_{i.j}}$.
Some expressions for $\mathcal{C}_{\mathrm{Sys.}}^{\mathrm{label}}$
are denoted as $X|Y$, the implication is the same as that in
Table~\ref{Csyslabel1}.\label{Csyslabel3}}
\setlength{\tabcolsep}{3.35mm} {
\begin{tabular}{c|cccccccc}
\hline\hline
$\mathcal{C}_{\mathrm{Sys.}}^{\mathrm{label}}$& $T_{i.1}$ &$T_{i.2}$&$T_{i.3}$&$T_{i.4}$&$B_{i.1}$&$B_{i.2}$&$B_{i.3}$&$B_{i.4}$\\
\hline
$\Xi_c^\prime\bar{D}^\ast$&$g^2$&$g^2$&$\frac{1}{4}g_1^2\big|\frac{1}{2}g_2^2$&$\frac{1}{4}g_3^2$&$\frac{3}{8}g^2g_1^2\big|\frac{3}{4}g^2g_2^2$&$\frac{3}{8}g^2g_1^2\big|\frac{3}{4}g^2g_2^2$&$\frac{3}{8}g^2g_3^2$&$\frac{3}{8}g^2g_3^2$\\
$\Xi_c\bar{D}^\ast$&$g^2$&$g^2$&$g_6^2\big|\frac{1}{2}g_2^2$&$\frac{1}{2}g_4^2$&$\frac{3}{2}g^2g_6^2\big|\frac{3}{4}g^2g_2^2$&$\frac{3}{2}g^2g_6^2\big|\frac{3}{4}g^2g_2^2$&$\frac{3}{4}g^2g_4^2$&$\frac{3}{4}g^2g_4^2$\\
$\Xi_c^\ast\bar{D}^\ast$&$g^2$&$g^2$&$\frac{1}{4}g_5^2$&$\frac{1}{4}g_3^2\big|\frac{1}{2}g_4^2$&$\frac{3}{8}g^2g_5^2$&$\frac{3}{8}g^2g_5^2$&$\frac{3}{32}g^2g_3^2\big|\frac{3}{16}g^2g_4^2$&$\frac{3}{32}g^2g_3^2\big|\frac{3}{16}g^2g_4^2$\\
\hline\hline
\end{tabular}
}
\end{table*}

\section{Possible molecular pentaquarks in $\Xi_c\bar{D}^{(\ast)}$, $\Xi_c^\prime\bar{D}^{(\ast)}$ and $\Xi_c^\ast\bar{D}^{(\ast)}$ systems}\label{NumericalResults}

In order to get the numerical results, we have to determine the
eight LECs in Eqs.~\eqref{Contact_Lag_BM}
and~\eqref{Contact_Lag_B3M}. As in Ref.~\cite{Meng:2019nzy}, we also
propose a SU(3) quark model to estimate the LECs. One can find the
detailed derivations in Appendix~\ref{app:QM}.

\begin{figure*}
\begin{center}
\begin{minipage}[t]{0.32\linewidth}
\centering
\includegraphics[width=\columnwidth]{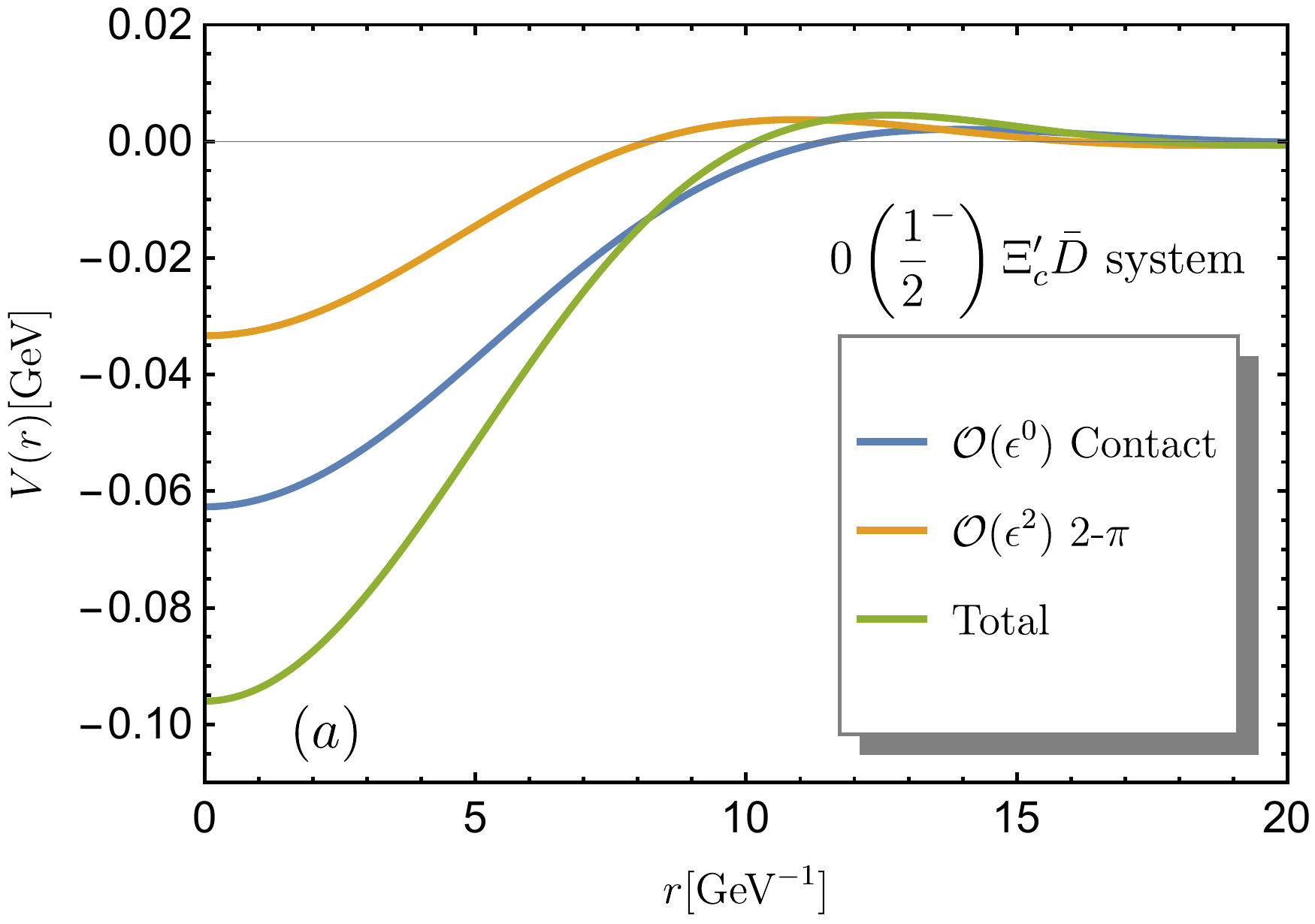}
\end{minipage}%
\begin{minipage}[t]{0.32\linewidth}
\centering
\includegraphics[width=\columnwidth]{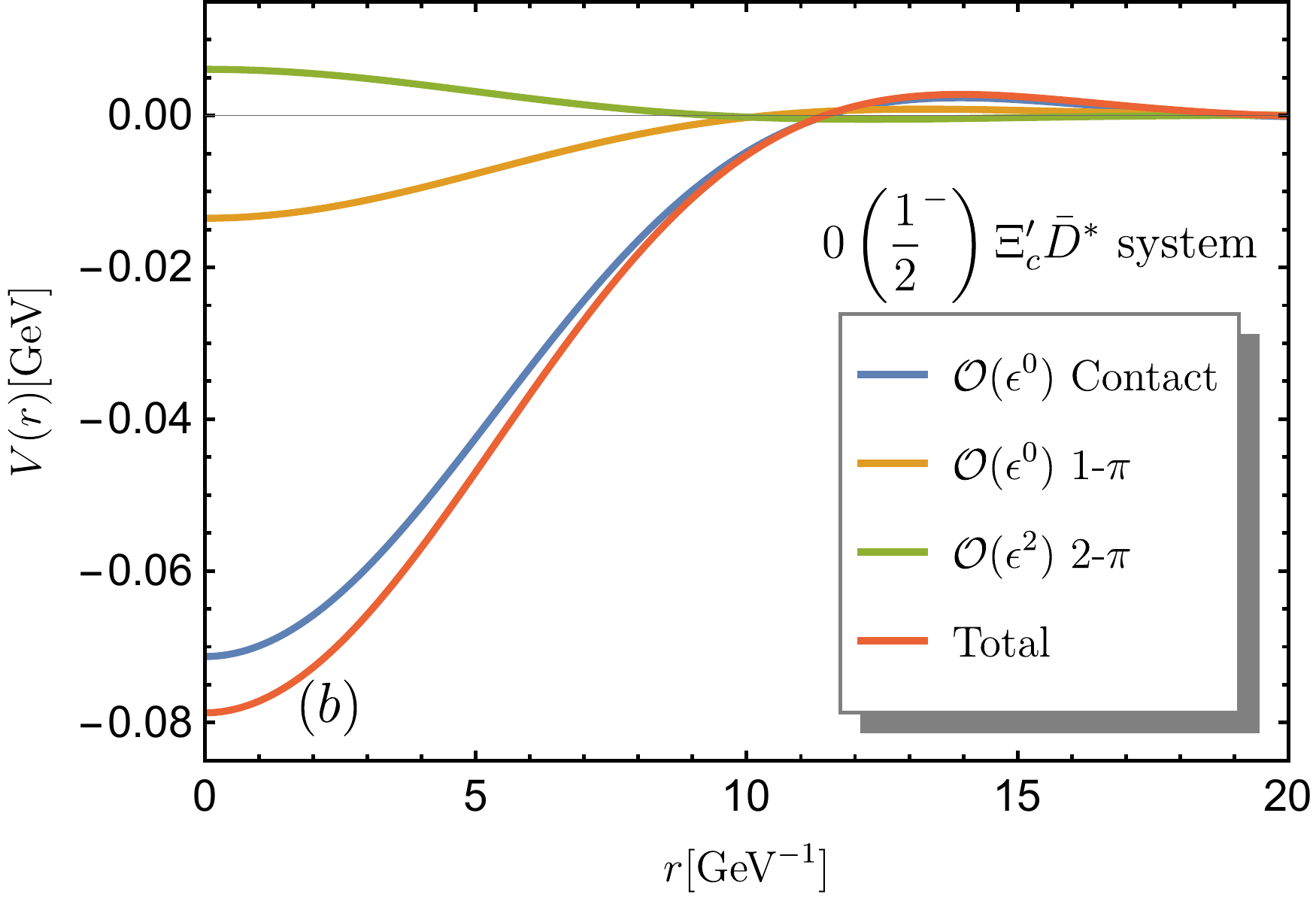}
\end{minipage}
\begin{minipage}[t]{0.32\linewidth}
\centering
\includegraphics[width=\columnwidth]{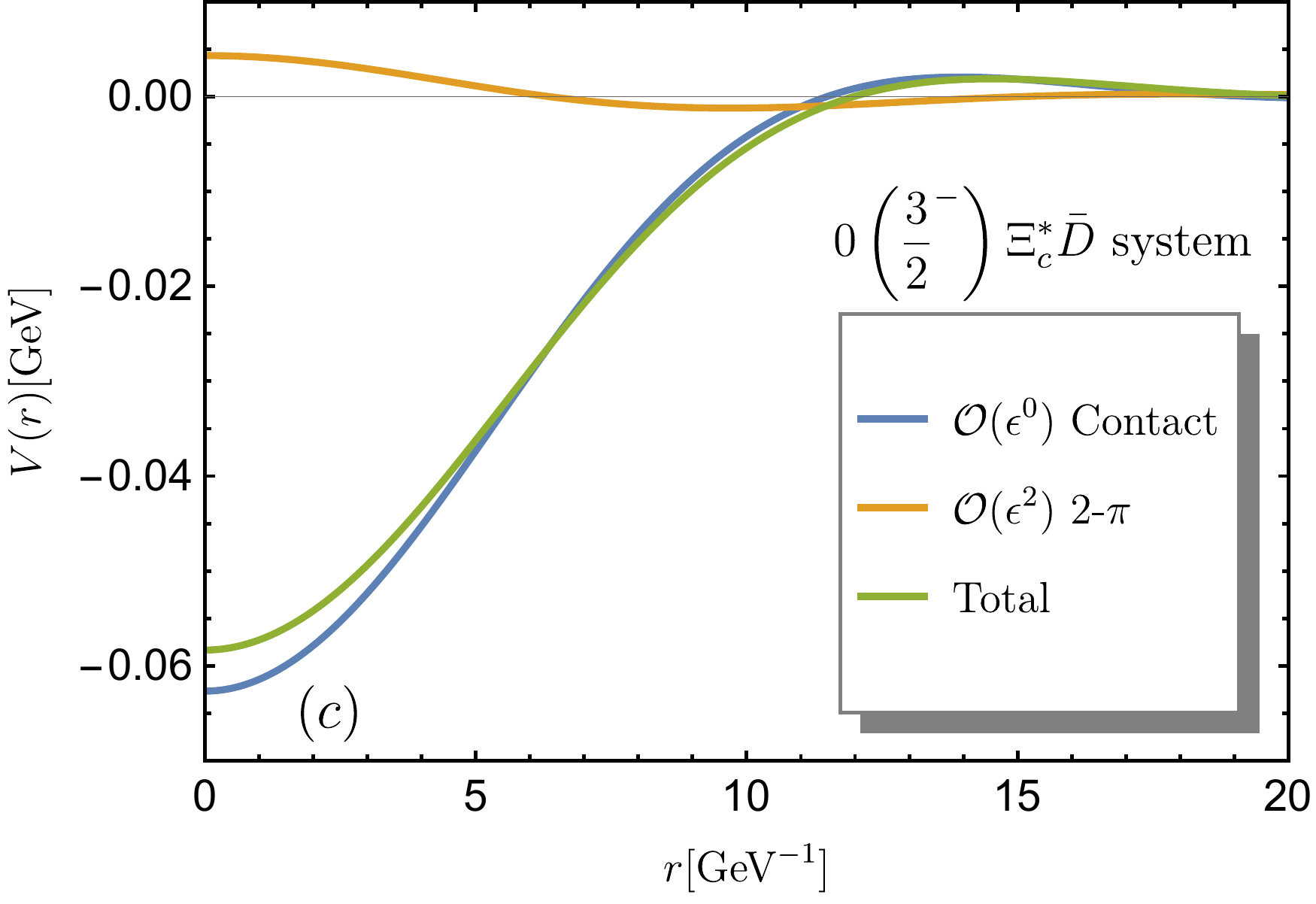}
\end{minipage}%
\\
\begin{minipage}[t]{0.32\linewidth}
\centering
\includegraphics[width=\columnwidth]{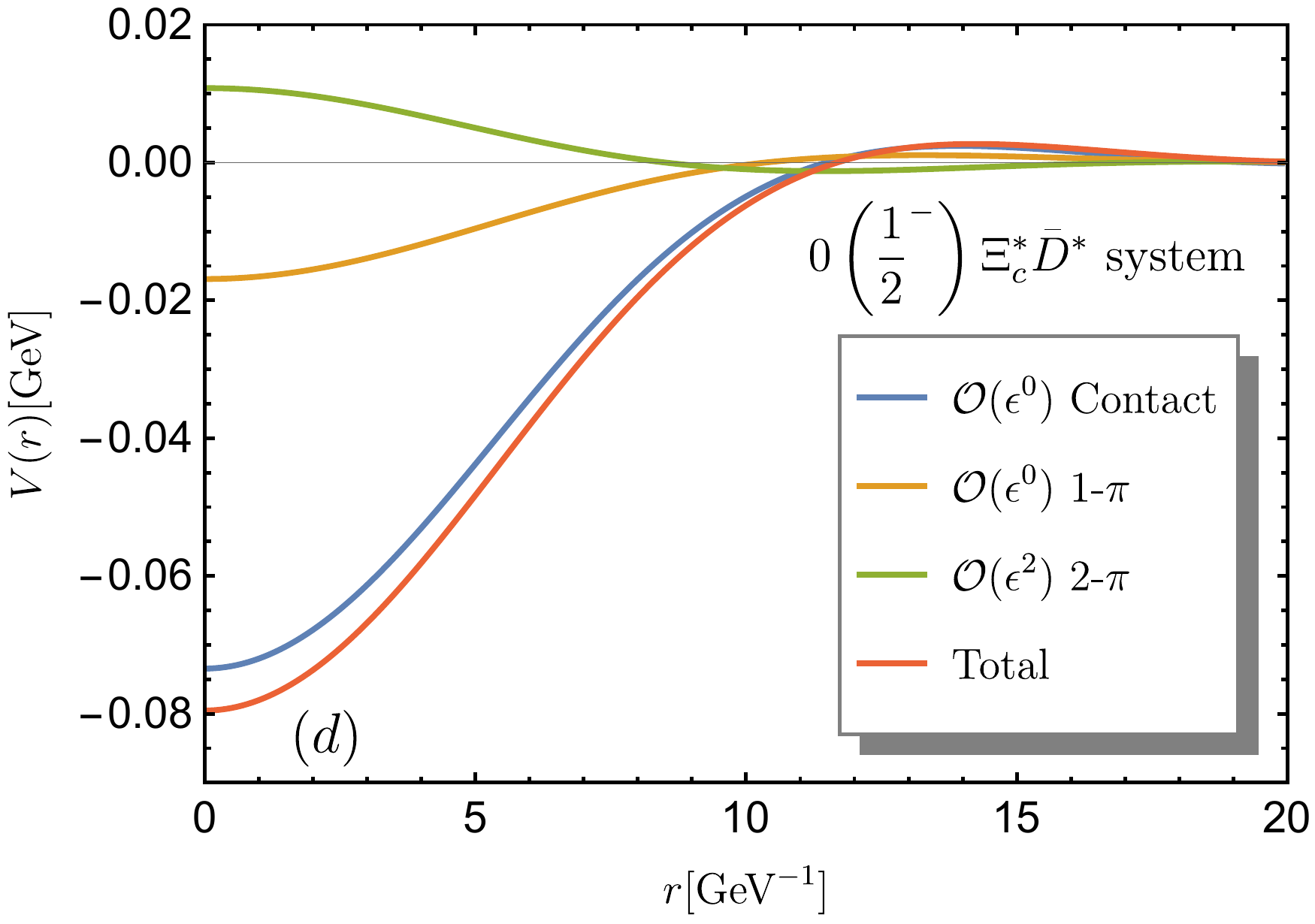}
\end{minipage}
\begin{minipage}[t]{0.32\linewidth}
\centering
\includegraphics[width=\columnwidth]{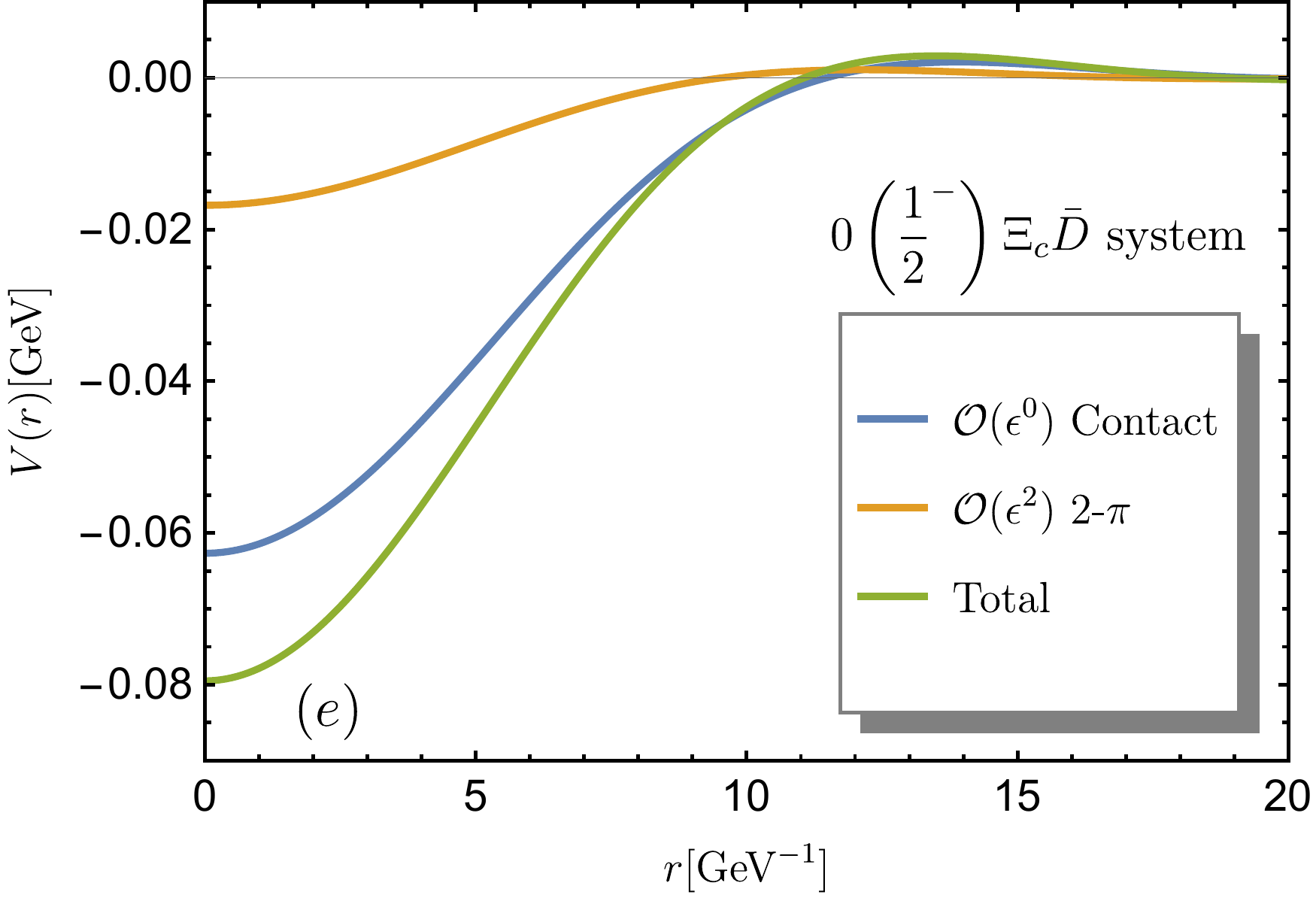}
\end{minipage}
\begin{minipage}[t]{0.32\linewidth}
\centering
\includegraphics[width=\columnwidth]{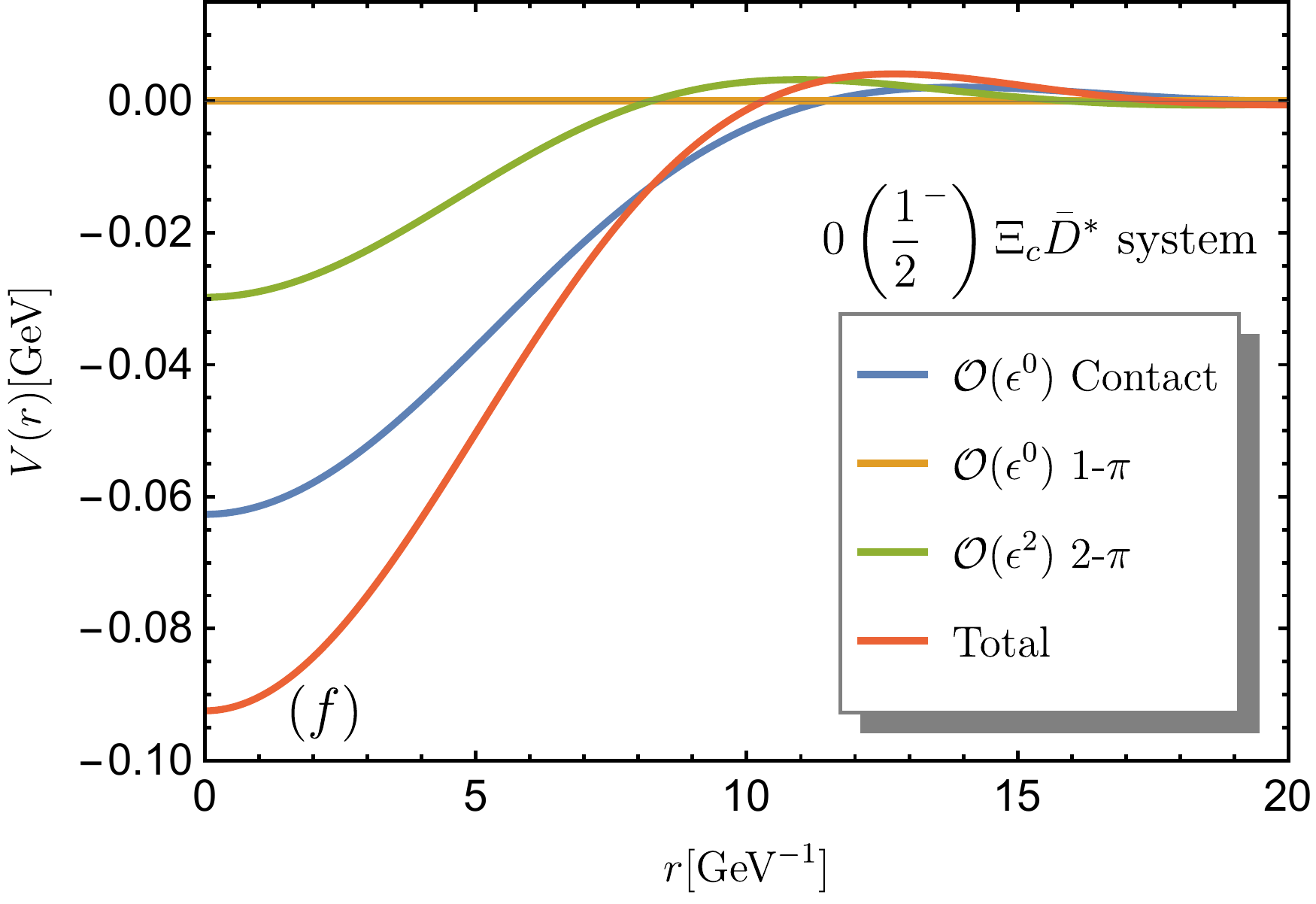}
\end{minipage}
\caption{The effective potentials of the
$\Xi_c^\prime\bar{D}^{(\ast)}$, $\Xi_c^\ast\bar{D}^{(\ast)}$ and
$\Xi_c\bar{D}^{(\ast)}$ systems. Their $I(J^P)$ are marked in each
subfigure. The potentials are obtained with the cutoff parameter
$\Lambda=0.4$ GeV. For the effective potentials of the unlisted spin
multiplets, their total potentials have the similar behaviors with
the lowest spin states.\label{Potential2}}
\end{center}
\end{figure*}

The effective potentials of some representative $I=0$ channels are
given in Fig.~\ref{Potential2}. We notice that the leading order
contact interaction supplies a strong attractive potential for all
the considered channels. This situation is the same as those of the
$\Sigma_c^{(\ast)}\bar{D}^{(\ast)}$ systems~\cite{Wang:2019ato}. For
the $\Xi_c^\prime\bar{D}$ and $\Xi_c\bar{D}^\ast$ systems, one can
find the two-pion-exchange contributions are also significant
because of the accidental degeneration between $\Xi_c^\prime\bar{D}$
and $\Xi_c\bar{D}^\ast$ in the loops. However, the behavior of the
two-pion-exchange potentials for $\Xi_c^\prime\bar{D}$ and
$\Sigma_c\bar{D}$ are totally different due to the opposite sign of
the mass differences. For example, the mass difference between
$\Sigma_c\bar{D}$ and $\Lambda_c\bar{D}^\ast$ is about $28$ MeV,
while that for $\Xi_c^\prime\bar{D}$ and $\Xi_c\bar{D}^\ast$ is
about $-32$ MeV. By solving the Schr\"odinger equation, we find the
bound states in the $\Xi_c^\prime\bar{D}^{(\ast)}$ and
$\Xi_c^\ast\bar{D}^{(\ast)}$ systems, likewise. The predicted
binding energies and masses are given in Table~\ref{Csyslabel2}.

Theoretically, the existence of the bound states in the
$\Xi_c^\prime\bar{D}^{(\ast)}$ and $\Xi_c^\ast\bar{D}^{(\ast)}$
systems is not a surprise, because $\Xi_c^\prime$ and $\Xi_c^\ast$
belong to the same flavor multiplets with the $\Sigma_c$ and
$\Sigma_c^\ast$, respectively in the SU(3) flavor
symmetry~\cite{Peng:2019wys}. Nevertheless, things become
interesting when we go to the $\Xi_c\bar{D}^{(\ast)}$ systems. On
the one hand, no bound states are experimentally observed near the
$\Lambda_c\bar{D}^{(\ast)}$ thresholds up to
now~\cite{Aaij:2019vzc}. On the other hand, some theoretical
calculations also do not support the existence of bound states in
$\Lambda_c\bar{D}^{(\ast)}$ systems~\cite{Yang:2011wz,Wang:2011rga}.
As the SU(3) partner of $\Lambda_c$, the interactions of the
$\Xi_c\bar{D}^{(\ast)}$ systems show totally different behavior at
the short range [e.g., see figures~\ref{Potential2}$(e)$
and~\ref{Potential2}$(f)$]. The attractive contact interaction is
strong enough to form bound states.

For the $\Lambda_c\bar{D}^{(\ast)}$ and $\Xi_c\bar{D}^{(\ast)}$
systems, the long-range one-pion-exchange vanishes, thus only the
contact term and two-pion-exchange contribute to their potentials.
The isospin $I$ and spin $J^\ell$ ($J^\ell$ denotes the total spin
of the light degrees of freedom in $\Lambda_c$) of the $\Lambda_c$
are both zero, so there are no isospin-isospin and spin-spin
interactions for the $\Lambda_c\bar{D}^{(\ast)}$ systems\footnote{At
the leading order of the heavy quark expansion, the spin-spin
interaction between $\Lambda_c$ and $\bar{D}^\ast$ are represented
by their light degrees of freedom, i.e.,
$\bm{J}^\ell_{\Lambda_c}\cdot \bm{J}^\ell_{\bar{D}^\ast}$. Its
matrix element vanishes for the $\Lambda_c\bar{D}^{\ast}$ system.}.
Ignoring the $\eta$ and $K$ meson contributions in the loops, one
can roughly get the two-pion-exchange potential of the
$\Lambda_c\bar{D}^{(\ast)}$ from the $\Xi_c\bar{D}^{(\ast)}$
expressions by setting the matrix element
$\langle\mathbf{I}_1\cdot\mathbf{I}_2\rangle$ to be zero. Although
the two-pion-exchange potential of the $\Lambda_c\bar{D}^{(\ast)}$
(or say the contribution from couple-channel effect) is attractive,
the one from the leading order $\tilde{D}_a$ term [e.g., see
Eq.~\eqref{Contact_Lag_B3M}] is repulsive. Their contributions
almost cancel each other, thus there are no binding solutions for
the $\Lambda_c\bar{D}^{(\ast)}$ systems.

The $J^\ell$ of $\Xi_c$ is also zero, thus there are no spin-spin
interactions for the $\Xi_c\bar{D}^{(\ast)}$ systems as well. But
$I=\frac{1}{2}$ for $\Xi_c$, i.e., the isospin-isospin interaction
in the $\Xi_c\bar{D}^{(\ast)}$ systems is the main reason that leads
to the different scenarios for the $\Lambda_c\bar{D}^{(\ast)}$ and
$\Xi_c\bar{D}^{(\ast)}$ systems. The leading order isospin related
$\tilde{E}_a$ term provides a very strong attractive potential, as
well as the attractive two-pion-exchange potential. Their
contributions together yield three bound states in the
$\Xi_c\bar{D}^{(\ast)}$ systems (see Table~\ref{Csyslabel2}). A
recent study based on the local hidden gauge approach also gives a
similar conclusion~\cite{Xiao:2019gjd}.

We also tried to include the one-eta-exchange contribution for the
$\Xi_c^\prime\bar{D}^\ast$ and $\Xi_c^\ast\bar{D}^\ast$ systems at
the leading order, where we use the experimental values of $f_\eta$
and $m_\eta$ as
inputs~\cite{Jiang:2015xqa,Meng:2018gan,Wang:2018cre,Wang:2019mhm}.
We notice the one-eta-exchange contribution is marginal, which only
introduces about $1\%$ and even much smaller than $1\%$ corrections
to the $\Xi_c^\prime\bar{D}^\ast$ and $\Xi_c^\ast\bar{D}^\ast$
binding energies, respectively.

Additionally, we also calculate the potentials of these six systems
in the $I=1$ channel. However, the total potentials are all
repulsive, i.e., no bound states exist for the
$\Xi_c\bar{D}^{(\ast)}$, $\Xi_c^\prime\bar{D}^{(\ast)}$ and
$\Xi_c^\ast\bar{D}^{(\ast)}$ systems in the isovector channels.

These $I=0$ molecular pentaquarks with a strange quark can be
reconstructed at the $J/\psi\Lambda$ final states. We hope LHCb
collaborations may search for these new $P_{cs}$s near the
$\Xi_c\bar{D}^{(\ast)}$ thresholds. Some discussions on the
$\Lambda_c\bar{D}_s^{(\ast)}$, $\Sigma_c\bar{D}_s^{(\ast)}$,
$\Sigma_c^\ast\bar{D}_s^{(\ast)}$ and $\Omega_c^{(\ast)}\bar{D}_s^{(\ast)}$ systems
are given in Appendix \ref{app:other}.

\begin{table*}[htbp]
\centering
\renewcommand{\arraystretch}{1.8}
\caption{The predicted binding energies $\Delta E$ and masses $M$
for the $[\Xi_c^\prime\bar{D}^{(\ast)}]_J$,
$[\Xi_c^\ast\bar{D}^{(\ast)}]_J$ and $[\Xi_c\bar{D}^{(\ast)}]_J$
systems in $I=0$ channel, where the subscript ``$J$" denotes the
total spin of the system. We correspondingly use the thresholds of
$\Xi_c^{\prime+}\bar{D}^{(\ast)0}$, $\Xi_c^{\ast+}\bar{D}^{(\ast)0}$
and $\Xi_c^+\bar{D}^{(\ast)0}$ as the benchmarks to calculate the
values in this table (in units of MeV). The state that denoted by
``$\sharp$" means which may be nonexistent at the upper
limit.\label{Csyslabel2}} \setlength{\tabcolsep}{1.05mm} {
\begin{tabular}{c|ccccccc|ccc}
\hline\hline
System&$[\Xi_c^\prime\bar{D}]_{\frac{1}{2}}$&$[\Xi_c^\prime\bar{D}^\ast]_{\frac{1}{2}}$&$[\Xi_c^\prime\bar{D}^\ast]_{\frac{3}{2}}$&$[\Xi_c^\ast\bar{D}]_{\frac{3}{2}}$&$[\Xi_c^\ast\bar{D}^\ast]_{\frac{1}{2}}$
&$[\Xi_c^\ast\bar{D}^\ast]_{\frac{3}{2}}$&$[\Xi_c^\ast\bar{D}^\ast]_{\frac{5}{2}}^\sharp$&$[\Xi_c\bar{D}]_{\frac{1}{2}}$&$[\Xi_c\bar{D}^\ast]_{\frac{1}{2}}$&$[\Xi_c\bar{D}^\ast]_{\frac{3}{2}}$\\
\hline
$\Delta E$&$-18.5^{+6.4}_{-6.8}$&$-15.6^{+6.4}_{-7.2}$&$-2.0^{+1.8}_{-3.3}$&$-7.5^{+4.2}_{-5.3}$&$-17.0^{+6.7}_{-7.5}$&$-8.0^{+4.5}_{-5.6}$&$-0.7^{+0.7}_{-2.2}$&$-13.3^{+2.8}_{-3.0}$&$-17.8^{+3.2}_{-3.3}$&$-11.8^{+2.8}_{-3.0}$\\
$M$&$4423.7^{+6.4}_{-6.8}$&$4568.7^{+6.4}_{-7.2}$&$4582.3^{+1.8}_{-3.3}$&$4502.9^{+4.2}_{-5.3}$&$4635.4^{+6.7}_{-7.5}$&$4644.4^{+4.5}_{-5.6}$&$4651.7^{+0.7}_{-2.2}$&$4319.4^{+2.8}_{-3.0}$&$4456.9^{+3.2}_{-3.3}$&$4463.0^{+2.8}_{-3.0}$\\
\hline\hline
\end{tabular}
}
\end{table*}

\section{Summary}\label{Summary}

In this work, we have systematically calculated the effective
potentials of $\Xi_c\bar{D}^{(\ast)}$,
$\Xi_c^\prime\bar{D}^{(\ast)}$ and $\Xi_c^\ast\bar{D}^{(\ast)}$
systems with the chiral effective field theory up to the
next-to-leading order. The contact interaction, one-pion-exchange
contribution and two-pion-exchange diagrams are considered. By
fitting the newly observed $P_c$ spectra, we relate the LECs to
those of the $\Sigma_c\bar{D}^\ast$ systems with the quark model
(see Appendix~\ref{app:QM}).

As the partners of $\Sigma_c\bar{D}^{(\ast)}$ and
$\Sigma_c^\ast\bar{D}^{(\ast)}$, we also find seven bound states in
the isoscalar $[\Xi_c^\prime\bar{D}^{(\ast)}]_J$ and
$[\Xi_c^\ast\bar{D}^{(\ast)}]_J$ systems. The contact terms provide
the attractive potentials, which are dominant for these systems. The
two-pion-exchange interactions are important to the
$\Xi_c^\prime\bar{D}$ and $\Xi_c\bar{D}^\ast$ systems due to the
accidental degeneration of the intermediate states in the loops.

With the estimated LECs, we also obtain three bound states in the
isoscalar $[\Xi_c\bar{D}^{(\ast)}]_J$ systems. This is very
different from their partners $\Lambda_c\bar{D}^{(\ast)}$. Our
analyses do not support the existence of any molecular pentaquarks
in the $\Lambda_c\bar{D}^{(\ast)}$ systems. The difference between
$\Lambda_c\bar{D}^{(\ast)}$ and $[\Xi_c\bar{D}^{(\ast)}]_J$ arises
from the isospin-isospin interaction, which vanishes for the
$\Lambda_c\bar{D}^{(\ast)}$.

We considered the influence of one-eta-exchange interaction. Its
contribution only gives rise to $1\%$ corrections to the magnitude
of the binding energies. Thus the tiny effect is neglected in our
numerical results. Our calculation indicates that the potentials of
the $I=1$ channels are all strongly repulsive and no bound states
exist in these $I=1$ channels.

In summary, we obtain ten molecular pentaquarks $P_{cs}$ in the
isoscalar $\Xi_c\bar{D}^{(\ast)}$, $\Xi_c^\prime\bar{D}^{(\ast)}$
and $\Xi_c^\ast\bar{D}^{(\ast)}$ systems. Their signals can be
reconstructed in the $J/\psi\Lambda$ final states at LHCb
experiment.

\section*{Acknowledgments}
B. Wang is very grateful to R. Chen, X. L. Chen and W. Z. Deng for
helpful discussions. This project is supported by the National
Natural Science Foundation of China under Grant 11975033.

\begin{appendix}

\section{Estimating the LECs with quark model}\label{app:QM}

In this part, we estimate the eight LECs in
Eqs.~\eqref{Contact_Lag_BM} and~\eqref{Contact_Lag_B3M} from the
viewpoints of quark model. We assume the short-range contact
interaction stems from some heavy particle exchanges, which is
analogous to the resonance saturation model~\cite{Epelbaum:2001fm}.
However, we do not specify the exchange particles (such as $\rho$,
$\omega$, $f_0$, etc..) as in the one-boson-exchange scheme, because
their contributions are partially mimicked by the two-pion-exchange
diagrams. Rather than calculating the contact potential at the
hadron level, we construct the quark-level Lagrangians to depict the
short-range interaction by borrowing some concepts from the
quark-hadron duality and quark model. Generally, one can formulate
the quark-level Lagrangians as follows,
\begin{eqnarray}\label{QLlAG}
\mathcal{L}=g_s\bar{q}\mathcal{S}q+g_a\bar{q}\gamma_\mu\gamma^5\mathcal{A}^\mu
q,
\end{eqnarray}
where $q=(u,d,s)$, $g_s$ and $g_a$ are two independent coupling
constants. The $\mathcal{S}$ and $\mathcal{A}^\mu$ are two
fictitious fields, which create (annihilate) the scalar and
axial-vector spurions respectively. We assume the $\mathcal{S}$ and
$\mathcal{A}^\mu$ are flavor octets and have the similar matrix form
as that in the second term of Eq.~\eqref{pifield}. They are
introduced to produce the central potential and spin-spin
interaction between two quarks, respectively.

With the quark-level Lagrangians, the $\Sigma_c\bar{D}^\ast$
effective potential can be expressed as
\begin{eqnarray}\label{Pquark}
V_{\Sigma_c\bar{D}^\ast}^{\mathrm{Q.L.}}&=&\langle\Sigma_c\bar{D}^\ast|\mathcal{L}|\Sigma_c\bar{D}^\ast\rangle=-\frac{g_s^2}{6m_{\mathcal{S}}^2}-\frac{g_s^2}{m_{\mathcal{S}}^2}(\mathbf{I}_1\cdot\mathbf{I}_2)\nonumber\\
&&+\frac{g_a^2}{9m_{\mathcal{A}}^2}\bm{\sigma}\cdot\mathbf{T}+\frac{2g_a^2}{3m_{\mathcal{A}}^2}(\bm{\sigma}\cdot\mathbf{T})(\mathbf{I}_1\cdot\mathbf{I}_2),
\end{eqnarray}
where the superscript ``Q.L." is the abbreviation of quark level. We
take the SU(3) flavor symmetry and ignore the mass differences in
the $\mathcal{S}$ and $\mathcal{A}^\mu$ multiplets, respectively. In
addition, the assumption that $q^2\ll
m_{\mathcal{S}}^2~(m_{\mathcal{A}}^2)$ is used. Thus if we know the
square of the ``charge-to-mass ratios" $g_s^2/m_{\mathcal{S}}^2$ and
$g_a^2/m_{\mathcal{A}}^2$, we could obtain the potentials of the
other systems that contain the light quarks. The contact potential
of the $\Sigma_c\bar{D}^\ast$ in the $I=\frac{1}{2}$ channel is
given as~\cite{Wang:2019ato}
\begin{eqnarray}\label{Phadron}
V_{\Sigma_c\bar{D}^\ast}&=&-\mathbb{D}_1-\frac{2}{3}\mathbb{D}_2(\bm{\sigma}\cdot\mathbf{T}).
\end{eqnarray}
By fitting the $P_c$ spectra with the cutoff $\Lambda=0.4$ GeV, we
get $\mathbb{D}_1=63.1$ GeV$^{-2}$, $\mathbb{D}_2=6.5$ GeV$^{-2}$.
Defining $C_1=g_s^2/m_{\mathcal{S}}^2$,
$C_2=g_a^2/m_{\mathcal{A}}^2$, and comparing Eq.~\eqref{Pquark} and
Eq.~\eqref{Phadron}, one easily gets
\begin{eqnarray}
C_1=-\frac{6}{5}\mathbb{D}_1,\qquad C_2=\frac{6}{5}\mathbb{D}_2.
\end{eqnarray}

Following the same procedure, we can also calculate the
$\Xi_c^\prime\bar{D}^\ast$ contact potential with the quark-level
Lagrangians, which yield
\begin{eqnarray}\label{PXicDastH}
V_{\Xi_c^\prime\bar{D}^\ast}^{\mathrm{Q.L.}}=\left[\frac{1}{12}-(\mathbf{I}_1\cdot\mathbf{I}_2)\right]\left(C_1-\frac{2}{3}C_2\bm{\sigma}\cdot\mathbf{T}\right).
\end{eqnarray}
Matching Eq.~\eqref{PXicDast} and Eq.~\eqref{PXicDastH}, one can
obtain the LECs in Eq.~\eqref{Contact_Lag_BM} under the SU(3)
symmetry, which read
\begin{eqnarray}
D_a=0,\quad D_b=0,\quad E_a=\frac{1}{2}C_1,\quad
E_b=-\frac{1}{2}C_2.
\end{eqnarray}

Similarly, one can also calculate the contact potential of
$\Xi_c\bar{D}^\ast$ at the quark level with the Lagrangians in
Eq.~\eqref{QLlAG},
\begin{eqnarray}\label{QLXicDast}
V_{\Xi_c\bar{D}^\ast}^{\mathrm{Q.L.}}=\left[\frac{1}{12}-(\mathbf{I}_1\cdot\mathbf{I}_2)\right]C_1.
\end{eqnarray}
A match between Eq.~\eqref{PpXicDast} and Eq.~\eqref{QLXicDast}
gives
\begin{eqnarray}
\tilde{D}_a=0,\quad \tilde{D}_b=0,\quad
\tilde{E}_a=-\frac{1}{4}C_1,\quad \tilde{E}_b=0,
\end{eqnarray}
where $\tilde{D}_b=\tilde{E}_b=0$ is the consequence of $J^\ell=0$
for the light diquark in $\Xi_c$, i.e., the spin-spin interaction
vanishes for the $\mathcal{B}_{\bar{3}}$ baryons.

Because the exchange particles we considered in Eq.~\eqref{QLlAG}
are only octets, thus the final results show that the LECs $D_a$,
$D_b$ and $\tilde{D}_a$ are all zero. Their values should be
contributed by the singlets exchange. We could estimate the $D_a$,
$D_b$ and $\tilde{D}_a$ by replacing the octets in $\mathcal{S}$ and
$\mathcal{A}$ with the nonet. An alternative way is to expand the
Gell-Mann matrices in Eqs.~\eqref{Contact_Lag_BM}
and~\eqref{Contact_Lag_B3M} with $\lambda_0=\sqrt{2/ 3}\text{
diag}\{1,1,1\}$. The extra $\lambda_0$ terms can be matched to the
$D_a$, $D_b$ and $\tilde{D}_a$ terms. The relations read
\begin{eqnarray}
D_a={2\over 3}E_a,\quad D_b={2\over 3} E_b,\quad \tilde{D}_a={2\over
3} \tilde{D}_a. \label{DaDbDat}
\end{eqnarray}

We attempt to include the influences of $D_a$, $D_b$ and
$\tilde{D}_a$ on the numerical results in the SU(3) case.
Considering the masses of the singlets are heavier than those of the
octets, we adopt the half values in Eq.~\eqref{DaDbDat} as their
limits to give a conservative estimation. Finally, the values of the
these LECs are given in Table~\ref{LECvalues}. 
\begin{table}[!htbp]
\centering
\renewcommand{\arraystretch}{1.3}
\caption{The values of the LECs in Eqs.~\eqref{Contact_Lag_BM}
and~\eqref{Contact_Lag_B3M} estimated with quark model (in units of
GeV$^{-2}$).\label{LECvalues}} \setlength{\tabcolsep}{6.25mm} {
\begin{tabular}{cccc}
\hline\hline
$D_a$&$D_b$&$E_a$&$E_b$\\
$0\pm12.6$&$0\pm1.3$&$-37.9$&$-3.9$\\
\hline
$\tilde{D}_a$&$\tilde{D}_b$&$\tilde{E}_a$&$\tilde{E}_b$\\
$0\pm3.2$&$0$&$18.9$&$0$\\
\hline\hline
\end{tabular}
}
\end{table}
\section{Some discussions on the $\Lambda_c\bar{D}_s^{(\ast)}$, $\Sigma_c\bar{D}_s^{(\ast)}$, $\Sigma_c^\ast\bar{D}_s^{(\ast)}$ and $\Omega_c^{(\ast)}\bar{D}_s^{(\ast)}$ systems}\label{app:other}

In addition to the $\Xi_c\bar{D}^{(\ast)}$,
$\Xi_c^\prime\bar{D}^{(\ast)}$ and $\Xi_c^\ast\bar{D}^{(\ast)}$
systems we presented above, the strange hidden charm configuration
can also be the $\Lambda_c\bar{D}_s^{(\ast)}$,
$\Sigma_c\bar{D}_s^{(\ast)}$ and $\Sigma_c^\ast\bar{D}_s^{(\ast)}$
systems, even the three strange systems
$\Omega_c^{(\ast)}\bar{D}_s^{(\ast)}$. In this appendix, we discuss
the possibility of the existence of bound states in these systems.
There are ten systems by unfolding the different combinations. We
first list their leading order contact potentials in the following,
\begin{eqnarray}
\mathcal{V}_{\Lambda_c\bar{D}_s}&=&2\tilde{D}_a-\frac{4}{3}\tilde{E}_a,\\
\mathcal{V}_{\Lambda_c\bar{D}_s^\ast}&=&2\tilde{D}_a-\frac{4}{3}\tilde{E}_a+\left[2\tilde{D}_b-\frac{4}{3}\tilde{E}_b\right]\boldsymbol{\sigma}\cdot\mathbf{T},\\
\mathcal{V}_{\Sigma_c\bar{D}_s}&=&-D_a+\frac{2}{3}E_a,\\
\mathcal{V}_{\Sigma_c\bar{D}_s^\ast}&=&-D_a+\frac{2}{3}E_a+\frac{2}{3}\left[-D_b+\frac{2}{3}E_b\right]\boldsymbol{\sigma}\cdot\mathbf{T},\\
\mathcal{V}_{\Sigma_c^\ast\bar{D}_s}&=&-D_a+\frac{2}{3}E_a,\\
\mathcal{V}_{\Sigma_c^\ast\bar{D}_s^\ast}&=&-D_a+\frac{2}{3}E_a+\left[-D_b+\frac{2}{3}E_b\right]\boldsymbol{\sigma}_{rs}\cdot\mathbf{T},\\
\mathcal{V}_{\Omega_c\bar{D}_s}&=&-D_a-\frac{4}{3}E_a,\\
\mathcal{V}_{\Omega_c\bar{D}_s^\ast}&=&-D_a-\frac{4}{3}E_a+\frac{2}{3}\left[-D_b-\frac{4}{3}E_b\right]\boldsymbol{\sigma}\cdot\mathbf{T},\\
\mathcal{V}_{\Omega_c^\ast\bar{D}_s}&=&-D_a-\frac{4}{3}E_a,\\
\mathcal{V}_{\Omega_c^\ast\bar{D}_s^\ast}&=&-D_a-\frac{4}{3}E_a+\left[-D_b-\frac{4}{3}E_b\right]\boldsymbol{\sigma}_{rs}\cdot\mathbf{T}.
\end{eqnarray}
Yet, there does not exist the one-pion-exchange potential for these
systems at the leading order. The exchange particle can only be the
$\eta$ meson for the $\Sigma_c\bar{D}_s^{(\ast)}$,
$\Sigma_c^\ast\bar{D}_s^{(\ast)}$ and
$\Omega_c^{(\ast)}\bar{D}_s^{(\ast)}$ systems. In the previous
sections, we notice that the influence of one-eta-exchange is rather
feeble. So the leading order contribution is dominantly from the
contact terms.

The situation becomes intractable to the next-to-leading order. On
the one hand, since the exchanged particles in the loops are the
kaon and $\eta$ meson, their masses are about four times larger than
the pion mass. The large mass of kaon and $\eta$ meson would make
the loop diagram contribution being immoderately enhanced, which
breaks the convergence of chiral expansion. In other words, the
prediction is not stable any more in this case. This is why we
did not consider the kaon and $\eta$ meson
exchange for the $\Xi_c\bar{D}^{(\ast)}$,
$\Xi_c^\prime\bar{D}^{(\ast)}$ and $\Xi_c^\ast\bar{D}^{(\ast)}$
systems in the loop diagrams. On the other hand, we obtained the
cutoff $\Lambda\simeq0.4$ GeV by fitting the newly observed $P_c$
states. The kaon and $\eta$ meson masses are obviously larger than
the cutoff, thus the kaon and $\eta$ meson contributions can be
regarded as being integrated out and packaged into the LECs.
Therefore, we do not try to explicitly calculate the two-kaon- and
two-eta-exchange contributions for these systems. Alternatively, we
focus on the contact interaction to conservatively investigate the
behavior of their potentials.

Using the estimated LECs in Table \ref{LECvalues}, the potentials of the
$\Lambda_c\bar{D}_s^{(\ast)}$, $\Sigma_c\bar{D}_s^{(\ast)}$ and
$\Sigma_c^\ast\bar{D}_s^{(\ast)}$ systems are attractive, but the
attractions are too weak to form the bound states. We also tried to
add half value to the LECs, but no binding solutions are found yet.
The potentials of the $\Omega_c^{(\ast)}\bar{D}_s^{(\ast)}$ systems
are all repulsive, thus there also are no bound states in the
$\Omega_c^{(\ast)}\bar{D}_s^{(\ast)}$ systems. The investigation on
the $\Xi_c\bar{D}^{(\ast)}$, $\Xi_c^\prime\bar{D}^{(\ast)}$ and
$\Xi_c^\ast\bar{D}^{(\ast)}$ systems shows that the strong
attractive potentials are provided by the isospin-isospin
interactions in the leading contact terms. However, for the
$\Lambda_c\bar{D}_s^{(\ast)}$, $\Sigma_c\bar{D}_s^{(\ast)}$,
$\Sigma_c^\ast\bar{D}_s^{(\ast)}$ and
$\Omega_c^{(\ast)}\bar{D}_s^{(\ast)}$ systems, there does not exist
the isospin-isospin interactions, thus our study indicates it is
hard to form the bound states in these systems.
\end{appendix}

\end{document}